\documentclass{iopart}

\usepackage{graphicx}% Include figure files
\usepackage{dcolumn}% eqnarray table columns on decimal point
\usepackage{mathrsfs}
\usepackage{iopams}
\usepackage{bm}
\usepackage{amssymb,amstext}
\usepackage{color,xspace}
\usepackage[small,compact,raggedright]{titlesec}
\usepackage{epstopdf}
\usepackage{soul,hyperref}

\newcommand{\ket}[1]{\left| {#1} \right\rangle}
\newcommand{\bra}[1]{\left\langle {#1} \right|}

\newcommand{\proj}[2]{\left| {#1} \right\rangle\!\left\langle {#2} \right|}

\newcommand{\eqref}[1]{(\ref{#1})}

\begin{document}

\title{Berry Phase Quantum Thermometer}
\author{Eduardo Mart\'{i}n-Mart\'{i}nez}
\address{Institute for Quantum Computing, University of Waterloo, 200 Univ.
Avenue W, Waterloo, Ontario N2L 3G1, Canada}
\address{Dept. of Applied Math. University of Waterloo, Ontario Canada N2L 3G1}
\address{Perimeter Institute for Theoretical Physics, Waterloo, Ontario N2L 2Y5, Canada}
\author{Andrzej Dragan}
\address{School of Mathematical Sciences, University of Nottingham, Nottingham NG7 2RD, United Kingdom}
\address{Institute of Theoretical Physics, University of Warsaw, Ho\.{z}a 69, 00-681 Warsaw, Poland}
\author{Robert B. Mann}
\address{Dept. Physics \& Astronomy, University of Waterloo, Ontario Canada N2L 3G1}
\address{Perimeter Institute for Theoretical Physics, Waterloo, Ontario N2L 2Y5, Canada}
\author{Ivette Fuentes \footnote{Published before under Fuentes-Guridi and Fuentes-Schuller}}
\address{School of Mathematical Sciences, University of Nottingham, Nottingham NG7 2RD, United Kingdom}

\begin{abstract}
We show how the Berry phase can be used to construct a high precision quantum thermometer. An important advantage of our scheme is that there is no need for the thermometer to acquire thermal equilibrium with the sample. This reduces measurement times and avoids precision limitations. 
\end{abstract}

\date{December 15th, 2011}


%\keywords{Entanglement, gravitational collapse, black holes, Hawking radiation}

\maketitle

\section{Introduction}
The state of a point-like discrete-energy level quantum system (e.g. an atom) interacting with a quantum field acquires a geometric phase \cite{Berryoriginal} that is dependent on the state of the field.   For pure field states, the phase encodes information about the number of particles in the field \cite{PhysRevLett.89.220404}. In particular, for initial squeezed states, the phase depends also on the squeezing strength \cite{PhysRevLett.85.5018}.  If the field is in a thermal state, this geometric phase encodes information about its temperature \cite{prl}, and so  was used in a proposal to measure the Unruh effect  at low accelerations \cite{prl}. 

In this article we use Berry phase to construct a high precision thermometer. The thermometer consists of an atomic interferometer and measures the temperature of a cold medium by comparison with a hotter thermal source of approximately known temperature. Since our scheme does not require the thermometer to reach thermal equilibrium with the sample, measurement times are shorter and precision limitations due to thermalization are avoided \cite{Limits}.

In principle the thermometer that we introduce can be physically implemented in different ways. Here, as a simple example, we consider the point-like system to be an atom whose internal level structure is described by a quantum harmonic oscillator. The sample will be a thermal source modeled as a quantum scalar field contained in a cavity. The sampled field could correspond to many possible different physical situations: a phonon field, an infinite number of harmonic oscillators coupled to the atom (a way to model thermal reservoirs), the electromagnetic field in one dimension or a fixed polarization component of a full 3D electromagnetic (EM) field. 

In the scenario we consider, the most favourable precision is obtained for temperatures of cold sources 3 orders of magnitude below the reference temperature of the hot source. Moreover, our thermometer is very sensitive to temperature variations of the cold sample but almost insensitive to variations of the hot source, making this setting a very precise and reliable method of measuring low temperatures with high precision. The temperature range in which the thermometer operates optimally is determined by the natural frequency of the harmonic oscillator (in this case the atom). Therefore, the thermometer will be tuned to a particular temperature range by choosing an atom with a given energy gap. It is possible to make measurements within different temperature ranges by using atoms with various energy gaps. If one employs atoms with energy gaps going from 1 Mhz to 1 Ghz, it is possible to measure with high precision temperatures over a range of $10^{-1}$ K to $10^{-6}$ K. Our proposal  is applicable in non-relativistic settings and is readily scalable using atoms with energy gaps of radio-frequency width to measure nanokelvin temperatures.

\section{Geometrical phase}

\begin{figure}[h]
\begin{center}
\includegraphics[width=1\textwidth]{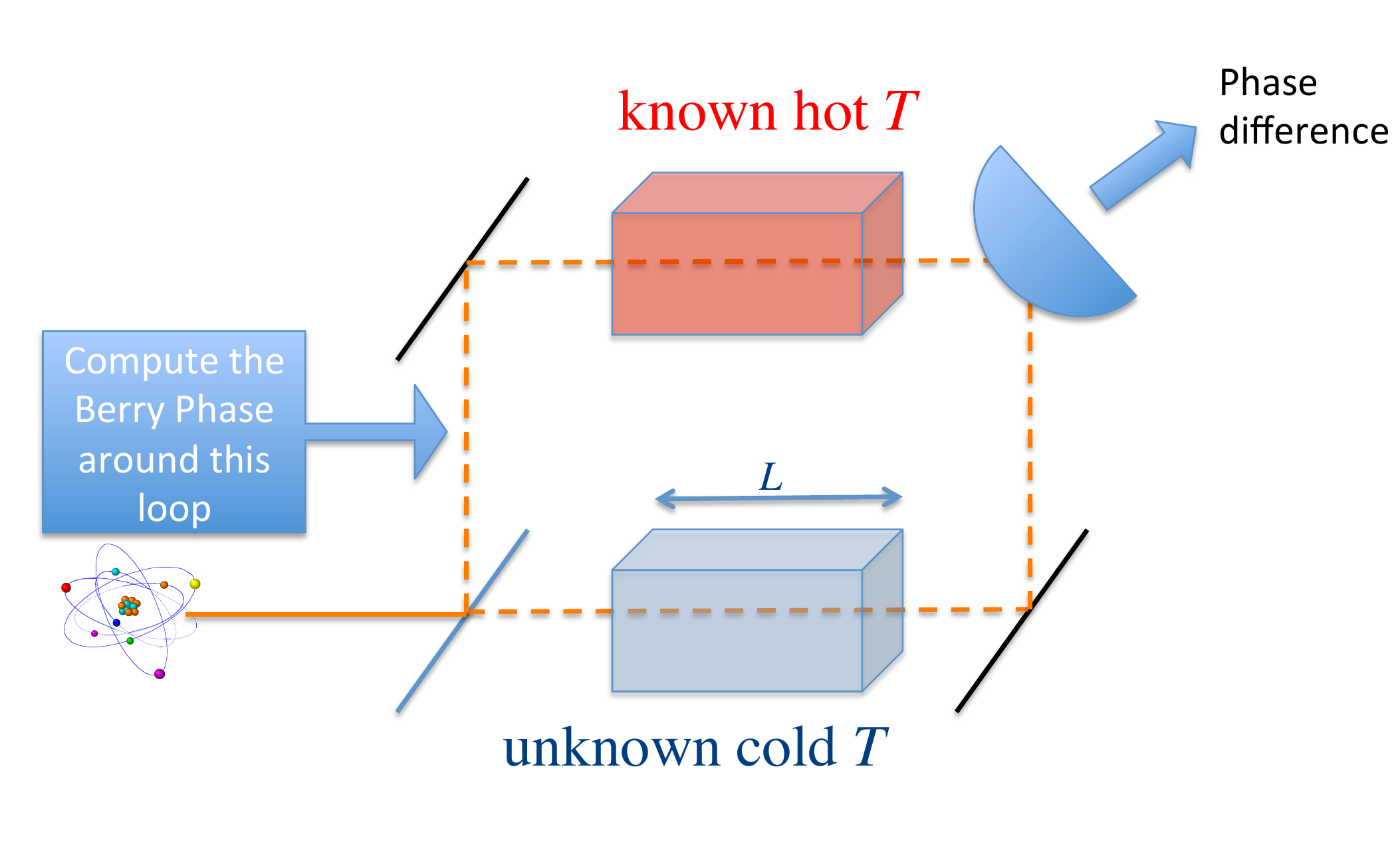}
\caption{Schematics of the setup: atomic interference of two atoms interacting with a  known hot source and an unknown temperature cold cavity.}
\label{fugu}
\end{center}
\end{figure}

To construct the Berry-phase-based thermometer consider an atomic interferometer in which a single atom follows two different paths (see Fig. \ref{fugu}). In one branch the atom goes through a cavity containing the cold sample while in the second branch the atom moves through a cavity where the field is in a thermal state of a known temperature. In both arms the atom will acquire a Berry phase due to the interaction with the field which will depend on the field's temperature. The geometric phase difference measured at the output of the interferometer will determine the temperature of the cold sample. 

We now proceed to  compute  the Berry phase acquired by an atom coupled to a quantum field inside a cavity.
 Let the atom to follow a trajectory given by $x(t)$ and its internal harmonic degrees of freedom be described by creation and annihilation operators  $d^{\dagger}$ and $d$. We consider the interaction Hamiltonian between the atom and the field to be given by:
\begin{equation}\label{hamilto}
H_\text{I}=\sum_m \lambda_m (d^\dagger e^{i\Omega t}+de^{-i\Omega t})(f_m^\dagger u_m(x(t))+\text{H.c.}).
\end{equation}
The $\lambda_m$'s are coupling constants, $\Omega$ and $t$ are the atom's frequency and proper time, respectively. The field operators $f_m^{\dagger}$ are associated with the field mode solutions $u_m(x(t))$ which are labeled by the frequency index $m$. These functions are evaluated at the atom's position implying that the interaction between the field and the atom takes place locally. For simplicity we are considering a $1+1$-dimensional field.

This model is a type of Unruh-DeWitt detector \cite{bigreview,DeWitt,LoSatz,LoSatz2,LoSch} that has been previously studied in other contexts \cite{prl,Massar,LinHu} and whose effectiveness and suitability has been deeply analyzed in \cite{UdWGauss,UdWsm1,UdWsm2}. It generally describes a multi-level quantum emitter (typically an atom) interacting with an infinite number of harmonic oscillators that may correspond to a scalar field. This model describes any harmonic detector coupled to the field (for example the electronic or vibrational spectrum of a many-level atom is a very well approximated by harmonic oscillators) 
and   is strongly related to the standard two-level approaches typically used to model atoms coupled to a scalar or EM field. In fact, there is a well-known mapping between this model and the standard Jaynes-Cummings model via the Holstein-Primakoff transformations  \cite{Holstein1}; the equivalence of the two models for the relevant regimes can be seen in \cite{Holstein2,Holstein3}.

Since we consider the field to be inside a cavity, the mode frequencies $m$ are discrete. For small cavities the mode separation is large and in this case it is possible to assume that the atom interacts effectively only with a single near-resonant cavity mode. In this case the analysis is considerably simplified and the Hamiltonian becomes
\begin{equation}\label{goodham2}
 H_I=\lambda(d^\dagger e^{i\Omega t}+de^{-i\Omega t})(f^\dagger u(x(t))+\text{H.c.}).
 \end{equation}
 and has been empirically verified \cite{Scullybook,Chenbook}. 
For a cavity of length $L$ with walls placed at $x=\pm(L/2)$, the normalized mode function is  given by
\begin{equation}\label{mode}
u(x(t))= \frac{1}{\sqrt{\omega L}}\sin\left(\omega[x+(L/2)]\right)e^{-i\omega t},
\end{equation}
where $\omega$ is the field frequency.   We assume the atom's trajectory is orthogonal to  $x$ such that  $x(t)=$const.  The effective coupling $\lambda$ seen by the atom would depend on the configuration of the cavity. We will work in the standard coupling regime $\lambda\approx 1-1.5$ Khz \cite{Scullybook}.

The most convenient procedure for computing the Berry phase is to operate in a mixed picture \cite{prl}, where the atom's free evolution is absorbed in the atoms' operators. This situation is mathematically more convenient for Berry phase calculations and the Hamiltonian (\ref{goodham2}) can be diagonalized analytically; its eigenstates are the dressed states $U^{\dagger}|m, n\rangle$, where $|m, n\rangle$ are eigenstates of the non-interacting Hamiltonian $H_0(\nu_f,\nu_d)=\frac12\nu_d\, d^\dagger d +\nu_f f^\dagger f$ and $U=S_f S_d  D_{fd}\hat S_d R_f$. The operators 
\begin{equation}
\begin{array}{ll}
\nonumber D_{fd}=\exp\big[s (f^\dagger d -  f d^\dagger)\big],&S_{f}=\exp\big[\frac12 u({f^\dagger}^2 - f^2)\big],\\*
\nonumber S_{d}=\exp\big[\frac12 v({d}^2 - {d^\dagger}^2) \big],&\hat S_{d}=\exp\big[p\, ({d}^2 - {d^\dagger}^2)\big]\\*
R_f=\exp\big(-i\varphi\, {f^\dagger f}\big),
\end{array}\label{sqops}
\end{equation}
are well-known in quantum optics and are the two-mode displacement, single-mode squeezing and phase rotation operators \cite{Scullybook}, respectively.

Obtaining the values of the parameters characterizing the squeezing, displacement and rotation, as functions of $(\lambda,\Omega,\omega)$ entails diagonalization of the Hamiltonian.  Since the eigenstates are $U^{\dagger}|m, n\rangle$, this is a straightforward
but tedious matter of carrying out basic algebraic manipulations, yielding  four independent parameters -- $v, \varphi, \nu_d$ and $\nu_f$ -- as  relatively complicated functions of $(\lambda,\Omega,\omega)$.  With these results it is now possible to calculate the Berry phase acquired by the atom due to its interaction with the field.

Berry showed that an eigenstate of a quantum system acquires a phase, in addition to the usual dynamical phase, when the parameters of the Hamiltonian are adiabatically and cyclicly varied with time \cite{Berryoriginal,aharonov,Amj}. In the case of a point-like atom interacting with a quantum field, the motion of the atom in spacetime gives rise to Berry's phase provided that the changes in the hamiltonian are adiabatic and cyclic. This phenomenon was recently exploited to propose an experiment to detect the Unruh effect with much smaller accelerations than previously considered \cite{prl}. Here we use it to build a quantum thermometer: when an atom interacts with a thermal state of a bosonic field, the geometric phase acquired is a function of the temperature of the probed state. 

The Berry phase acquired by the eigenstate $\ket{\psi(t)}$ of a system, whose Hamiltonian depends on $k$ cyclicly, is obtained by adiabatically varying the vector of parameters: $\bm R =(R_1(t),\dots,R_k(t))$:
\begin{equation}\label{Berry}
i\gamma=\oint_R\, \bm A \cdot  \text{d}\bm R,
\end{equation}
where $\bm{R}(t)$ is a closed trajectory in the parameter space and
\begin{equation}
\bm A= \left(
\bra{\psi(t)}\partial_{R_1}\ket{\psi(t)},
\ldots ,
 \bra{\psi(t)}\partial_{R_k}\ket{\psi(t)}\right).
 \end{equation}
 
As a first step, we calculate the Berry phase acquired by an eigenstate of the Hamiltonian under cyclic and adiabatic evolution of the parameters  $(v,\varphi,\nu_f,\nu_d)$.  The only non-zero component of $\bm A$ is 
$A_\varphi= \bra{0, n}S_aS_bD_{ab}R_a\partial_{\varphi} ( R_f^\dagger D^\dagger_{ab} S_d^\dagger S_f^\dagger)\ket{0, n}$, and so 
%\begin{eqnarray}i\gamma_I= \oint_{\varphi\in[0,2\pi)}\!\!\!\!\!\!\!\!\!\!\!\!\!\!\! \bm A \cdot  \text{d}\bm R=\int_{0}^{2\pi}\text{d}\varphi\, A_{\varphi}.\end{eqnarray}
 the Berry phase acquired by an eigenstate $U^\dagger\ket{0, n}$ is
\begin{equation}
\gamma_{I_{n}}=2\pi\bigg[\frac{\nu_d\, n \sinh(2v)\cosh[2(C-v)]}{\nu_f\sinh[2(C-v)]+\nu_d\sinh(2v)}+\gamma_{I_{0}}\bigg].
\end{equation}
where $C=\frac{1}{2}\ln\left(\nu_f/\nu_d\right)$ with  $\nu_f/\nu_d>e^{2v}$ and
\begin{equation}
\gamma_{I_{0}}=\frac{\nu_f \sinh^2 v \sinh [2(C-v)]+\nu_d\,\sinh(2v)\sinh^2 (C-v)}{\nu_f\, \sinh[2(C-v)]+\nu_d\,\sinh (2v)}.
\end{equation}

 Notice that the time dependence of the Hamiltonian that gives rise to the Berry phase is given by the time dependence of the parameter $\varphi$. In other words it is the movement of the detectors through space-time which makes the Hamiltonian time dependence cyclic, even if their position remains constant throughout the process, given the dependence of the field modes on time \eqref{mode}.

\section{Quantum thermometer} 

  We are now interested in calculating the Berry phase acquired by the atom when it interacts with a field in a  thermal state. The density matrix of a thermal state with temperature $T$ is given by \cite{prl}, 
\[\rho_{T}=\bigotimes_{\omega}\frac{1}{\cosh^2 r_{_T} }\sum_{n_\omega} \tanh^{2n} r_{_T} \proj{n_\omega}{n_\omega}\]
where
\[\tanh r_{_T} = \exp\left(-\frac{\hbar  \omega}{2k_\text{B}T}\right).\]

  We assume that, before the atom-field interaction is turned on, the system is in the mixed state  $\proj{0}{0}\otimes\rho_{T}$. If the interaction is then turned on adiabatically the state evolves into   $\rho= U^\dagger \left({\proj{0}{0}}\otimes\rho_T\right)U$.
 
 The geometric phase $\gamma = \text{Re}(\eta)$ acquired by a mixed state $\rho=\sum_i \omega_i \proj{i}{i}$ after a cyclic and adiabatic evolution  is given by $e^{i\eta} = \sum_i\omega_ie^{i\gamma_i}$ \cite{Vlatko} , where  $\gamma_i$ the Berry phase acquired by an eigenstate $\ket{i}$ of the Hamiltonian. We find that after a single cycle the state $\rho_T$ acquires the phase 
 \begin{equation}
 e^{i\eta} =\frac{e^{i\gamma_{I_{0}}}}{\cosh^2 r_{_T}-e^{2\pi\,i G}\sinh^2 r_{_T}}, 
\end{equation}
where
\begin{equation}
G= \frac{\nu_d\,  \sinh(2v)\cosh[2(C-v)]}{\nu_f\sinh[2(C-v)]+\nu_d\sinh(2v)}.
\end{equation}
Therefore  the Berry phase accrued by the state of the atom interacting with a thermal state is
\begin{equation}
\gamma_{_T}=\text{Re}\eta=\gamma_{I_{0}}-\text{Arg}\left(\cosh^2 r_{_T} - e^{2\pi\,i G } \sinh^2 r_{_T}\right).
\end{equation}
Let us compare the Berry phase acquired by the same multi-level system interacting with two thermal sources at different temperatures. After a complete cycle in the parameter space (near-resonance this means waiting a time $2\pi\Omega^{-1}$) the phase difference between these two atoms is equal to:
\begin{equation}\label{deltai}
\delta=\text{Arg}\left(1- e^{-\frac{\hslash\omega}{k_{\text{B}} T_1}-2\pi\,i G } \right)-\text{Arg}\left(1 - e^{-\frac{\hslash\omega}{k_{\text{B}} T_2}-2\pi\,i G } \right)
\end{equation}

\begin{figure}[h]
\begin{center}
\includegraphics[width=.70\textwidth]{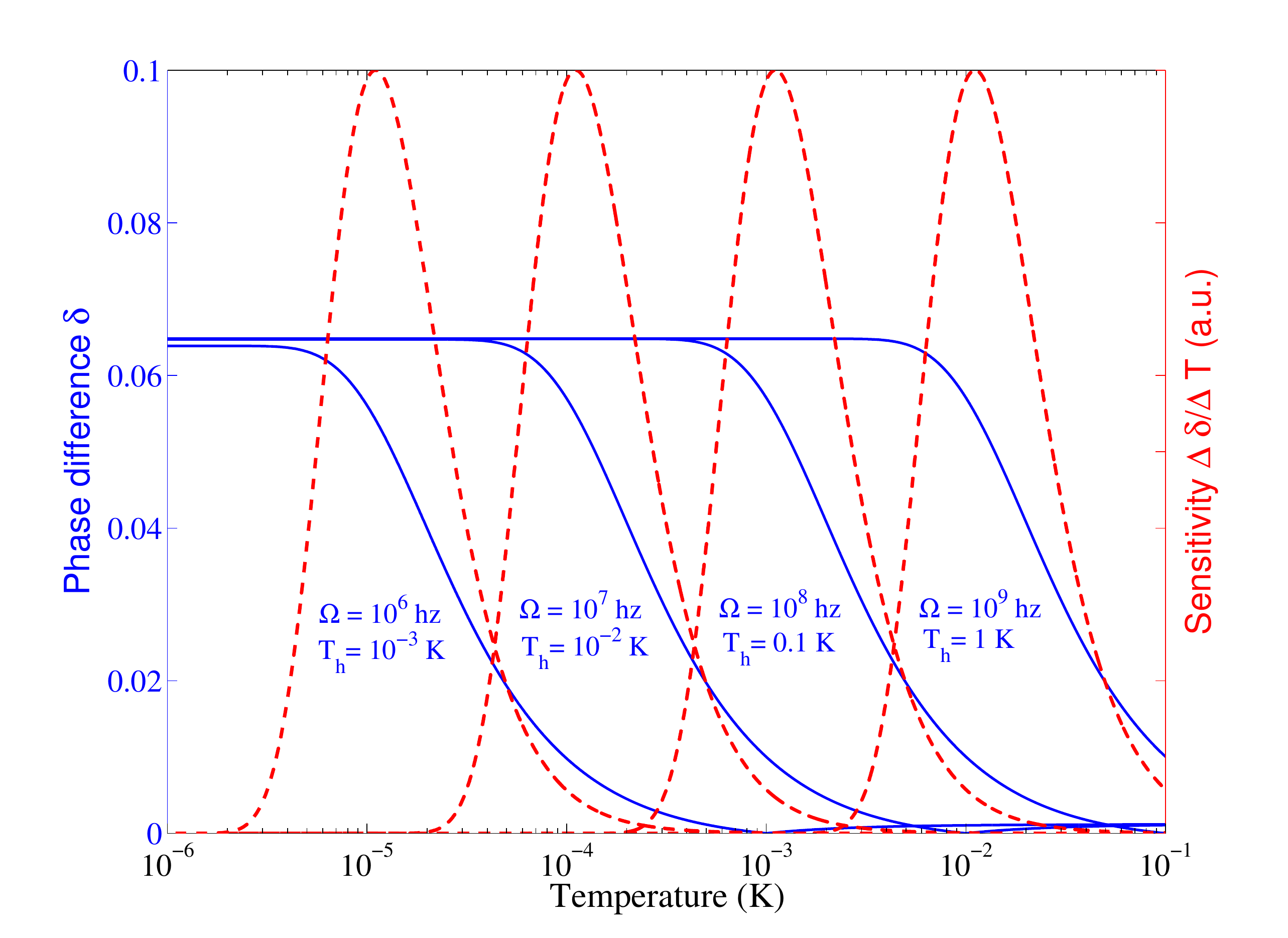}
\caption{Blue  solid lines: Geometric  phase difference $\delta$ between  2 detectors interacting with a cold and hot source of temperatures $T_{\text{c}}$ and $T_{\text{h}}$ respectively, as a function of the cold source temperature for different values of the atom gap and   hot source temperature. From left to right: $\Omega=10^6$ hz, $T_{\text{h}}=1$ mK; $\Omega=10^7$ hz, $T_{\text{h}}=10$ mK; $\Omega=10^8$ hz, $T_{\text{h}}=0.1$ K; $\Omega=10^9$ hz, $T_{\text{h}}=1$ K. Coupling frequency: 1.2 Khz for all the cases. Red dashed lines: Sensitivity curves for all the cases previously considered.}
\label{realf1}
\end{center}
\end{figure}

\begin{figure}[h]
\begin{center}
\includegraphics[width=.70\textwidth]{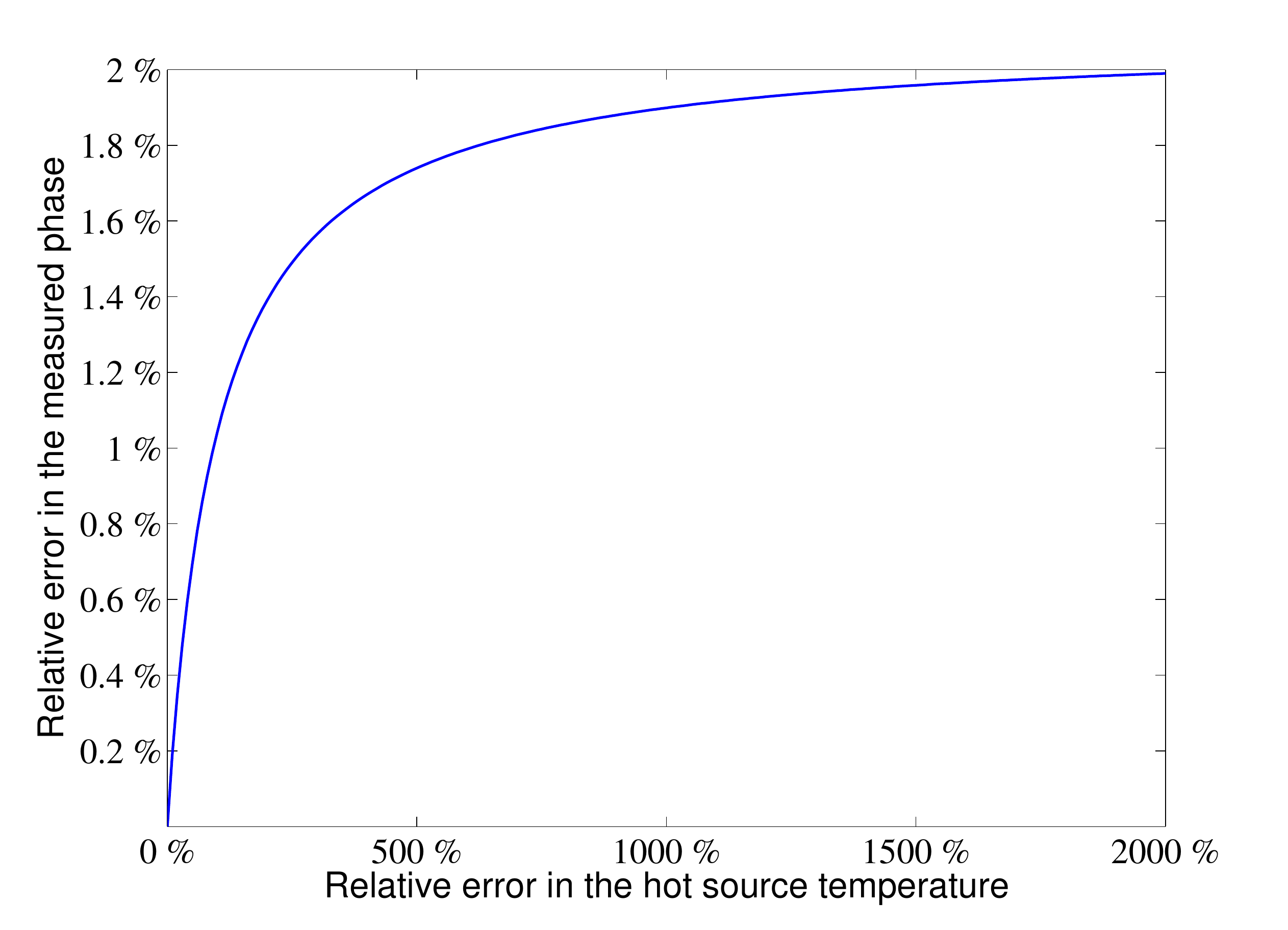}
\caption{Relative error in the Berry phase (and therefore, the determination of the temperature for the cold source) as a function of the relative error in determining the temperature of the hot source. As we see, the setting is very robust: huge changes of temperature of the hot source translate into small changes in the phase $\delta$.}
\label{precision}
\end{center}
\end{figure}

For realistic coupling values for atoms in cavities the phase difference $\delta$ is of considerable magnitude as shown in
Fig.~\ref{realf1}; it is also very sensitive to a particular range of temperatures, depending on the atomic gap. Adjusting this gap, we tune $\delta$ to a particular  temperature range, where we assume the hot source to have the temperature 3 orders of magnitude greater than the temperature of the target source.
While $\delta$ is quite sensitive to variations of the cold source, it is rather insensitive to changes in the hot source. Consequently we can obtain ultra-high precision measurements of the temperature of the cold source with almost no need to control the temperature of the hot one. In Fig. \ref{precision} we illustrate the effects of underestimating the temperature of the hot source on the precision of the measurement of $\delta$. Large variations of the hot source temperature translate into very small variations of the measured phase, providing us with an ultra-high precision thermometer, able to measure temperatures 3 orders of magnitude smaller than the hot source, over which no high precision temperature control is required.

The Berry phase is always a global phase. In order to detect it, it is necessary to prepare an interferometric experiment.  For example, a multi-level system in a superposition of  trajectories that go through different thermal sources would allow for detection of the phase. Any experimental set-up in which such a superposition can be implemented would serve our purposes. A possible scenario can be found in the context of atomic interferometry. This technology has already been successfully employed to measure with great precision general relativistic effects such as time dilation due to Earth's gravitational field \cite{atomGR}. Paths (of slightly different length) can be chosen such that the dynamical relative phase cancels. It is sufficient that the dynamical phase difference throughout both trajectories be equal or smaller than the geometric phase to allow for its detection. 

Although such cancellation depends upon the specific experimental setup, we find that even for a simple setting with current length metrology technology\footnote{$\text{Laserscale}^{\text{\textregistered}}$,
http://www.gebotech.de/pdf/\\LaserscaleGeneralCatalog\_en\_2010\_04.pdf}, we can control the relative dynamical phase with a precision several orders of magnitude smaller than the Berry phase acquired in one cycle.  

Consider an atom moving with speed  $\mathrm{v}$  along a trajectory of length $L$ in the interferometer. Changing the path length with a precision of  $\Delta L$ translates into changing the time of flight $T=L/\mathrm{v}$ with a precision  $\Delta T = \left|\frac{\partial T}{\partial L}\Delta L\right|=\frac{\Delta L}{\mathrm{v}}.$ Since the dynamical phase  $\phi\propto \Omega T$, 
the resulting precision is  approximately  $\Delta \phi=\Omega \Delta T\approx \Omega \Delta T=\frac{\Omega \Delta L}{\mathrm{v}}.$ For
 $\mathrm{v}\approx 100$ m/s, $\Omega\approx 2$ GHz, and  $\Delta L\approx 1 \cdot 10^{-11}$  m (see [20]) we obtain $\Delta \phi=2\cdot10^{-4}$.

\section{Possible limitations of the results} 

One might expect the atoms to lose coherence upon interacting with thermal sources, thus rendering the interferometry experiment useless.    However, for the energy gaps and temperatures considered here, to have decoherence times comparable with the time for one cycle the thermal sources  must be at temperatures  several orders of magnitude  above the ones we wish to measure. Furthermore, 
%there is a stronger condition that we need to fulfil, implying that the system is not decohered by the source: 
for this formalism to be valid we require that the probability of finding the atom in an excited state after one cycle of evolution should be much smaller than 1 (weak adiabadicity), in turn implying coherence is not lost.  In general we expect this condition to fail in three regimes: small atomic gap, high temperatures or high coupling. When it holds, the interaction time of the multi-level atom  with the thermal state should be short enough so that the only change in the atom's state is that it acquires a global phase (dynamical + geometrical).

To investigate under what regimes the time evolution of the coupled field-atom system fulfills this weak adiabaticity condition, we check that  the ground state of the atom for the Hamiltonian $H(t_0)$ evolves after a time  $t-t_0$ (equal to one cycle of evolution) to the ground state of the Hamiltonian $H(t)$  when the field is in a thermal state\footnote{Note that for realistic values of the coupling, the energy gap between the ground state and the first excitation of the interaction Hamiltonian is almost equal (different by approximately one part per million) to the free Hamiltonian.} \cite{prl}, a fact easily demonstrated for realistic values of the coupling $\lambda$.

Solving the exact  Schwinger equation (in the interaction picture):
\begin{equation}
\frac{d}{dt}\rho=-i\left[H_\text{I},\rho\right].
\end{equation}
numerically\footnote{Numerical computations were carried out with Matlab differential equation solvers.}, we find that indeed, after a short period of time (approximately $10^{4}\cdot 2\pi \Omega^{-1}$ for realistic couplings), the probability of finding the atom in the excited state cannot be distinguished from thermal noise, as expected  since for  short times a ground-state atom
in a thermal bath transiting to an excited state will start a Rabi-like oscillation, but for longer times the atom will thermalize. 
%\begin{figure}[h]
%\begin{center}
%\includegraphics[width=.70\textwidth]{Noise1e9L1e3}
%\includegraphics[width=.70\textwidth]{Noise1e9L1e6}
%\caption{Probability of excitation of the  of gap $\Omega=1$ Ghz (initially in the ground state) While interacting with a thermal bath of temperature 0.1 K. for different values of the coupling strength. Thermal noise is observed for the long times limit. the stronger the coupling the sooner the atom reaches the themal noise regime (top) $g=1$ khz [realistic regime]  (bottom) $g=1$ Mhz [ultra-strong coupling]]}
%\label{tremal}
%\end{center}
%\end{figure}

\begin{figure}[h]
\begin{center}
\includegraphics[width=.70\textwidth]{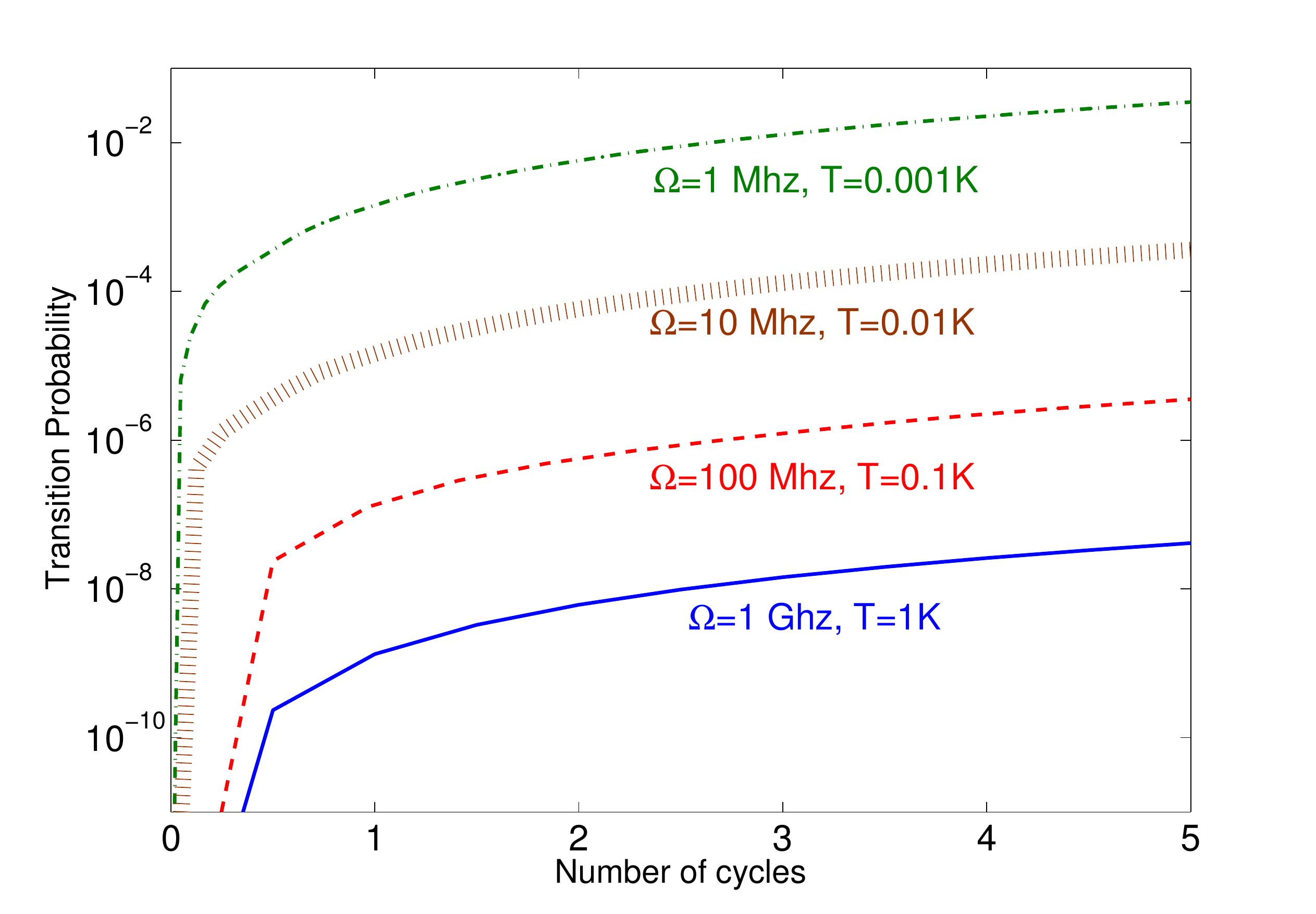}
\caption{Probability of atomic excitation in a time of a small number of cycles. (1 cycle means $t= 2\pi/\omega$ s). For the 1Ghz case $P<10^{-9}$, for the Mhz case $P<10^{-3}$}
\label{appcheck}
\end{center}
\end{figure}

However, for our purposes, we will only have the atoms interacting  with the thermal bath for very short times (1 cycle of evolution $t\approx 2\pi \Omega^{-1}$), and for these times we can see that for realistic values, the hypothesis holds perfectly. In Fig. \ref{appcheck} we can see that for the cases considered, even in the worst case scenario (1 Mhz gap and 1 mK temperature) the probability of excitation is $P\approx 10^{-3}\ll 1$. In the best case scenario considered (1 Ghz gap and 1 K temperature) the probability is $P\approx 10^{-9}\ll 1$  and the approximation holds very well.
In other words, the atoms interact with the thermal bath for a time short enough to avoid being excited; the only effect of the interaction is that they  acquire a phase dependent on the bath's temperature.

Furthermore, the probability remains negligible for many evolution cycles. Even in the worst-case-scenario considered, with realistic couplings and for an atomic gap of 1 Ghz, the approximation remains valid for more than 100 cycles until thermal fluctuations render the hypothesis invalid.
%of probability reach the off-hypothesis limit.
Considering one cycle of evolution, we need to demand that the transition probability $P\ll 1$. For the scenarios considered we find for the best case scenario $P\approx 10^{-9}$ (Ghz gap), while for the worst case scenario $P\approx 10^{-3}$ (Mhz gap).

As a final remark, the thermometer has a target temperature range for which it is highly sensitive. As seen in Fig.  \ref{realf1}, this range is typically of 2-3 orders of magnitude. To see an interference pattern in an atomic interference experiment that allows measurement of the target temperature, we need also a reference source at a fixed temperature for which we need to have some control but whose temperature may experience variations. To satisfy the necessary condition for reducing the influence of fluctuations in the reference source, in a manner shown in Fig. \ref{precision}, we need the temperature of the source to be 3 orders of magnitude different from the target temperature. In principle, the source could be hotter or colder than the target , but of course, it is much easier to control warmer thermal sources than colder ones.   As a final note, geometric phases can be generalized to  non-adiabatic and non-cyclic cases. We are currently considering generalizations of the thermometer in these directions.

\section{Acknowledgements} We would like to thank M. Berry, J. Le\'on, M. Montero and T. Ralph for their comments and suggestions.  I.~F. and A. ~D. thank EPSRC  [CAF Grant EP/G00496X/2] for financial support, A.~D. was supported by the Polish Ministry of Education grant. R.B.M. was supported in part by  NSERC and E. M-M would like to acknowledge the support of the NSERC Banting postdoctoral fellowship programme.  

\appendix

\section{Diagonalization of the Hamiltonian}

In this section we will show how to diagonalize a single-mode Unruh-Dewitt-like Hamiltonian describing the interaction of a harmonic oscillator with a scalar field. We will do it in a way which will be very convenient in order to deal with the Berry phase calculations.

We consider a point-like detector endowed with an internal structure which couples linearly to a scalar field $\phi(x(t))$ at a point $x(t)$ corresponding to the world line of the detector.
The interaction Hamiltonian is of the form $H_I\propto  \hat X \hat \phi(x(t))$ where 
we have chosen the detector to be modeled by a  harmonic oscillator with frequency $\Omega$. In this case the operator $\hat X\propto (d^\dagger + d)$ corresponds to the detector's position where $b^{\dagger}$ and $b$ are creation and anihilation operators .
Considering that the detector couples only to a single mode of the field with frequency  $|k|=\omega$, the field operator takes the form $\hat\phi(x(t))\approx\hat\phi_k(x(t))\propto \left[f\, e^{i(kx-\omega t)}+f^\dagger\, e^{-i(kx-\omega t)}\right]$, where $a^{\dagger}$ and $a$ are creation and anihilation operators associated to the field mode $k$.  The Hamiltonian is therefore
 \[H_T\!=\!\omega f^\dagger f+\Omega d^\dagger d + \lambda (d+d^\dagger)[f^\dagger e^{i(kx-\omega t)}+ fe^{-i(kx-\omega t)}].\]
 where $\lambda$ is the coupling frequency.
This Hamiltonian resembles the Unruh-DeWitt detector in the case where the atom interacts with a single mode of the field.  Standard calculations involving the Unruh-DeWitt detector employ the interaction picture since this is the most convenient picture to calculate transition probabilities. However, we employ a mixed picture where the detector's operators are time independent since this situation is mathematically more convenient for Berry phase calculations.

The Hamiltonian can be diagonalized analytically.  The eigenstates are given by $U^{\dagger}|N_a N_b\rangle$ where $U=S_a(u,\theta_a)  S_b(v,\theta_b)  D_{ab}(s,\phi)\hat S_b(p) R_a(\varphi) $ and $|N_a N_b\rangle$ are eigenstates of the diagonal Hamiltonian  $H_0(\omega_a,\omega_b)=\omega_a\, f^\dagger f+\omega_b\, d^\dagger d$ which determines the energy spectrum of the system. The operators are  slightly more general versions of those given in eq. (\ref{sqops})
\begin{eqnarray}
\nonumber D_{ab}=&D_{ab}(\chi)=\exp\big(\chi f^\dagger d - \chi^* f d^\dagger\big)\\*
\nonumber S_a=&S_{a}(\alpha)=\exp\big(\alpha^* {f^\dagger}^2 - \alpha f^2\big)\\*
\nonumber S_b=&S_{b}(\beta)=\exp\big(\beta^* {d^\dagger}^2 - \beta d^2\big)\\*
\nonumber  \hat S_b=&\hat S_{b}(p)=\exp\big[p\, ({d^\dagger}^2 - d^2)\big] \\*
R_a=&R_{a}(\varphi)=\exp\big(-i\varphi\, {f^\dagger f}\big)
\end{eqnarray}
where $\chi=s\,e^{i\phi}$, $\alpha=\frac12u\, e^{i\theta_a}$, $\beta=\frac12 v\,e^{i\theta_b}$. 
In order to satisfy the equation
\[H_{T}(\omega_a,\omega_b,\alpha,\beta,\chi,\varphi)=U^{\dagger}H_0U\]
the following constraints must be satisfied
\begin{eqnarray*}\label{constr}
s=\arctan\sqrt{\frac{\omega_a\,\sinh 2u}{\omega_b\,\sinh 2v}},\;\, \phi=\theta_a\!=  0,\;\,\theta_b=\!\pi,\;\, u=C-v
\end{eqnarray*}
where $C=\frac{1}{2}\ln\left(\omega_a/\omega_b\right)$ with  $\omega_a/\omega_b>e^{2v}$.
The dependence of the Hamiltonian parameters on  the parameters in the unitaries which diagonalize it are given by
\begin{eqnarray}\nonumber\omega=&\frac{\sinh 2v\left[\cosh \left[2(C-v)\right] +\frac{\sinh\left[2(C-v)\right]}{\tanh 2v}\right]}{{\omega_a^{-1}}\sinh 2v+\omega_b^{-1}\,\sinh \left[2(C-v)\right]}\\[3mm]
\nonumber\Omega = &\sqrt{\hat\Omega^2-4Z^2}\\[3mm]
\nonumber\lambda=&e^{p 
}\frac{\sqrt{\omega_a\omega_b\,\sinh [2(C-v)]\,\sinh 2v}}{\omega_b\,\sinh 2v+\omega_a\,\sinh [2(C-v)]}
\left[\omega_a\,\cosh \left[2(C-v)\right] - \omega_b\cosh 2v\right]\\[3mm]
\varphi = &kx-\omega t
\end{eqnarray}
where $2p=\rm{arctanh}\big[-2Z/\hat \Omega\big]$ and
\begin{eqnarray}
\nonumber\hat\Omega&=& \frac{\sinh 2v\left[\omega_a^2\, \frac{\sinh \left[4\left(C-v\right)\right]}{2\sinh 2v}+\omega_b^2\,\cosh 2v\right]}{\omega_b\,\sinh 2v+\omega_a\,\sinh [2(C-v)]}  \left[\omega_a\,\cosh \left[2(C-u)\right] - \omega_b\cosh 2u\right]\\[3mm]
Z&=&\frac12 \frac{\sinh 2v\left(\omega^2_a\frac{\sinh^2 \left[2(C-v)\right]}{\sinh 2v}- \omega_b^2 \sinh 2v\right)}{\omega_b\,\sinh 2v+\omega_a\,\sinh [2(C-v)]}
\end{eqnarray}

\vspace{1cm}
\bibliographystyle{bibstyleNCM}
\bibliography{references}

\end{document}